\pdfoutput=1
\documentclass[11pt]{article}

\usepackage[final]{acl}

\usepackage{bm}

\usepackage{times}
\usepackage{latexsym}
\usepackage{bbm}
\usepackage{amsmath}
\usepackage{mathtools}
\usepackage{algorithm}
\usepackage{algorithmic}
\usepackage{threeparttable}
\usepackage{multirow}
\usepackage{threeparttable}
\usepackage{amsmath}
\usepackage{bbm}
\usepackage{amsmath}
\usepackage{mathtools}
\usepackage{amsthm}
\usepackage{algorithm}
\usepackage{algorithmic}
\usepackage{amsmath}
\usepackage{threeparttable}
\usepackage{booktabs}
\usepackage{amsfonts}
\usepackage{subcaption}
\usepackage{xcolor}
\usepackage{amsmath}
\usepackage{threeparttable}
\usepackage[table]{xcolor}
\usepackage{tcolorbox}
\tcbuselibrary{skins, breakable} % <--- 必须加上这一行

\usepackage{url}
\usepackage{enumitem} % 控制列表间距

\usepackage{booktabs}
\usepackage{multirow}
\usepackage{graphicx}

\usepackage[T1]{fontenc}
\usepackage[utf8]{inputenc}

\usepackage{microtype}

\usepackage{inconsolata}

\usepackage{graphicx}

\title{A Unified Framework for Modeling Heterogeneous Financial Data via Dual-Granularity Prompting}

\author{
\textbf{Yu Lei} \textsuperscript{1,2},
\textbf{Zixuan Wang} \textsuperscript{1}\thanks{\ Corresponding Authors.},
\textbf{Yiqing Feng} \textsuperscript{1,2}, 
\textbf{Junru Zhang} \textsuperscript{3}, 
\textbf{Yahui Li} \textsuperscript{2} \\
\textbf{Chu Liu} \textsuperscript{1}, 
\textbf{Tongyao Wang} \textsuperscript{1}, 
\textbf{Dongyang Li} \textsuperscript{1} 
\\[5pt]
\textsuperscript{1} DiDi International Business Group\\ 
\textsuperscript{2} Beijing University of Posts and Telecommunications\quad \textsuperscript{3} Zhejiang University
}

\begin{document}
\maketitle

\begin{abstract}
Recent industrial credit scoring models remain heavily reliant on manually tuned statistical learning methods. Despite their potential, deep learning architectures have struggled to consistently outperform traditional statistical models in industrial credit scoring, largely due to the complexity of heterogeneous financial data and the challenge of modeling evolving creditworthiness. To bridge this gap, we introduce FinLangNet, a novel framework that reformulates credit scoring as a multi-scale sequential learning problem. FinLangNet processes heterogeneous financial data through a dual-module architecture that combines tabular feature extraction with temporal sequence modeling, generating probability distributions of users' future financial behaviors across multiple time horizons. A key innovation is our dual-prompt mechanism within the sequential module, which introduces learnable prompts operating at both feature-level granularity for capturing fine-grained temporal patterns and user-level granularity for aggregating holistic risk profiles. Notably, real world deployment yielded a 6.3 pp improvement in KS, along with a 9.9\% reduction in bad debt rate.

\end{abstract}

\section{Introduction}
Credit risk prediction is a cornerstone for financial institutions to devise effective lending policies and informed decisions evaluating the solvency of borrowers \cite{genovesi2023standardizing}. This process is critical in minimizing loan default risks, which is essential for preserving low bad debt levels and mitigating financial losses in the multi-billion dollar credit industry \cite{cheng2020contagious}. Credit risk models perform binary classifications to discern good from bad customers, improving overdue risk prediction, and effectively managing bad debt while maintaining profitability \cite{zhao2023fintech}.

In industrial credit risk assessment, user data typically originates from multiple heterogeneous sources including credit reports, transaction histories, and product usage behaviors \cite{lu2023profit}. These multi-source data present significant challenges: they are inherently noisy, high-dimensional, sparse, and discrete, often with substantial missing values \cite{elia2023investigating}. Furthermore, user creditworthiness is not static but evolves dynamically over time, requiring models to capture both short-term behavioral changes and long-term credit profile evolution \cite{niazkar2024applications}.

The risk control industry predominantly relies on XGBoost for its stability and interpretability when handling irregular multisource financial data \cite{li2022hybrid}. The standard practice involves extensive feature engineering to create derivative features, followed by feature selection and XGBoost modeling \cite{song2023loan}. While this approach achieves strong performance, it suffers from critical limitations: 
(1) extensive feature engineering is time-consuming and requires substantial domain expertise; 
(2) static models fail to capture temporal dependencies in sequential user behaviors; 
(3) point-in-time predictions cannot model the dynamic evolution of creditworthiness across different time horizons.

Recent advancements in time-series \cite{yao2022multivariate}, sequential models \cite{yousefi2022adoption}, and graph models \cite{xue2024risk} have shown promise in capturing temporal dynamics. However, these methods still focus on point-in-time predictions, failing to model how user creditworthiness evolves across multiple future time horizons. While XGBoost performs well for immediate risk assessment, its effectiveness deteriorates for longer-term predictions. In practical scenarios, understanding user behavior across various future time windows is crucial for comprehensive risk management. Therefore, we reformulate credit scoring as a multi-scale behavioral representation learning problem, where the model learns to characterize user profiles across different future periods.

In this work, we propose FinLangNet, a novel framework that reformulates credit scoring as a multi-scale sequential learning problem. Drawing inspiration from transformer architectures' success in capturing long-range dependencies, FinLangNet processes heterogeneous financial data through a dual-module approach: (1) a non-sequential module for extracting high-order interactions from tabular features, and (2) a sequential module with an innovative dual-prompt mechanism. This mechanism introduces feature-level prompts for capturing fine-grained temporal patterns and user-level prompts for aggregating holistic risk profiles. Through multi-scale predictions with dynamically weighted loss functions, FinLangNet generates comprehensive representations that capture both static characteristics and evolving behavioral dynamics.

Our contributions are summarized as follows:
\begin{enumerate}[topsep=-2pt, partopsep=0pt, parsep=0pt, itemsep=2pt]
    \item We reformulate credit scoring from classification to dynamic multi-scale forecasting, enabling the model to generate future behavioral distributions that capture evolving creditworthiness patterns across different time horizons.

    \item We propose FinLangNet \footnote{\url{https://github.com/didiglobal-fintech-credit-risk/FinLangNet}}, a hybrid architecture with two complementary modules: a DeepFM-based non-sequential module for tabular data and a sequential module featuring an innovative dual-prompt mechanism. The dual-prompt design operates at feature-level and user-level granularities, enabling comprehensive representation learning from heterogeneous financial data.

    \item We achieve state-of-the-art results on multiple benchmarks. On the public UEA time series classification benchmark, FinLangNet outperforms existing methods by substantial margins, demonstrating its superiority as a general-purpose sequential modeling framework. The model also shows consistent improvements across various financial risk assessment metrics in real-world datasets.
    
    \item We deployed FinLangNet in our finance platform. It achieved a {6.3 pp absolute gain in KS} and {9.9\% relative reduction in bad debt rate} compared to the previous XGBoost-based system, demonstrating both superior risk discrimination capability and substantial financial impact in real-world industrial settings.
\end{enumerate}

\section{Preliminaries}
\label{sec:preliminaries}

Our goal is to learn a representation for predicting user credit risk by leveraging heterogeneous user data. We formulate the input as \( X = (m, z) \), consisting of two modalities. The non-sequential component \(m \in \mathbb{R}^M\) represents static attributes (e.g., user profiles) where \(M\) is the feature dimension. The sequential component \(z = \{z_1, \dots, z_S\}\) captures dynamic behaviors from \(S\) different sources (e.g., query records, loan logs). Each source \(z_s \in \mathbb{R}^{C_s \times T_s}\) contains multivariate quantitative features with channel size \(C_s\) and sequence length \(T_s\), which may vary across sources. To enable sequential processing, we process all temporal sources in chronological order. Visual illustrations and detailed specifications of these inputs are provided in Appendix \ref{sec:appendix_data}.

The prediction task is defined as learning a function \(f_{\theta} : X \rightarrow [0, 1]^L\). The output is a probability distribution over \(L\) different time scales, where each entry represents the likelihood of overdue behavior within a specific future horizon. The model is trained under supervision using ground-truth labels that indicate whether a default event occurred at the corresponding time scale, guiding the model to capture risk dynamics over time.

\section{Methodology}

\begin{figure*}[t]
\begin{center}
	\centerline{\includegraphics[width=0.99\textwidth,height=0.50\textwidth]{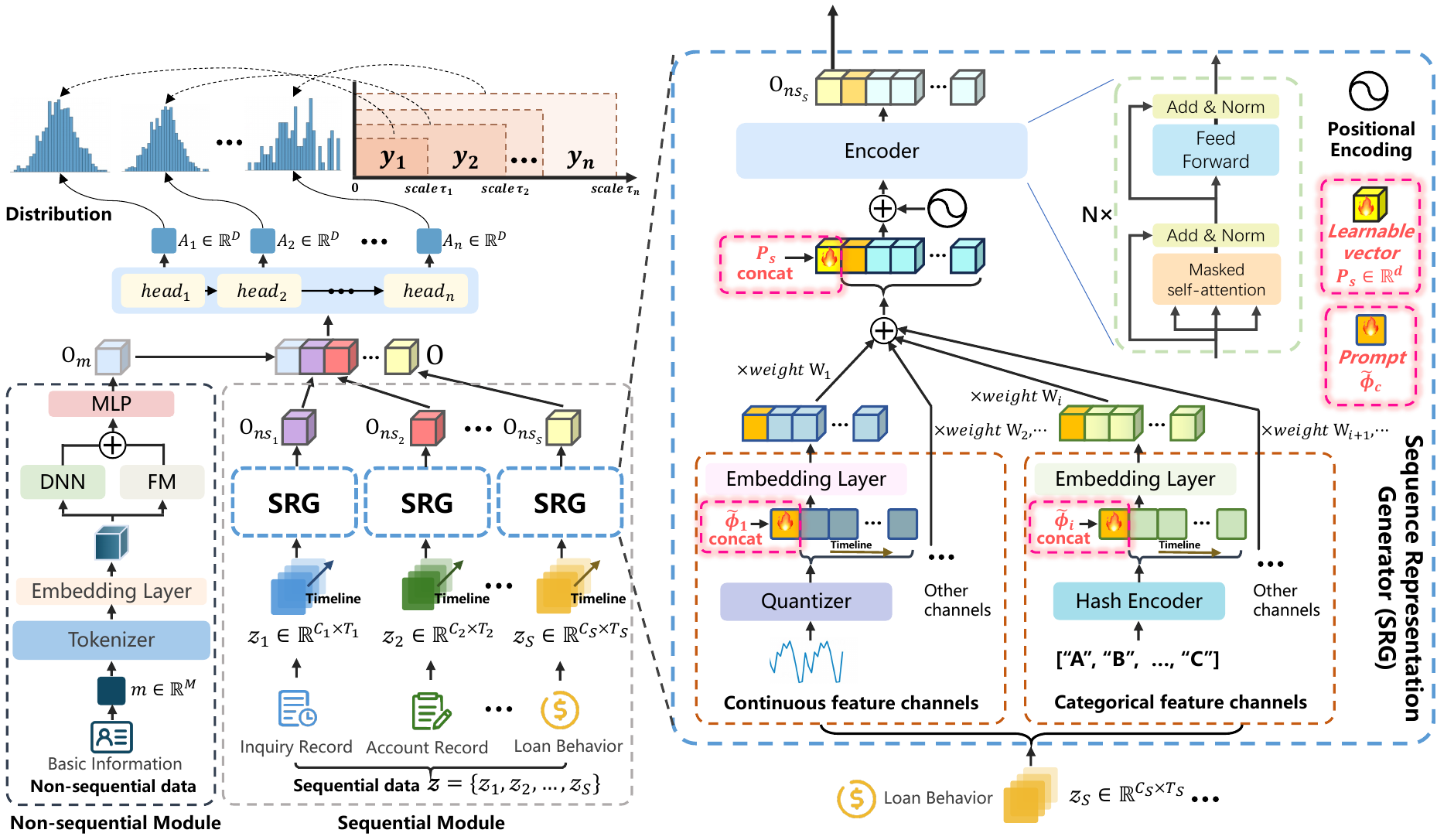}}
	\caption{Overview of the FinLangNet Framework. It processes static data via a Non-Sequential Module and temporal data via a Sequence Representation Generator (SRG). The SRG utilizes a dual-prompt mechanism to capture both feature-level and user-level patterns.}
	\label{fig:finlangnet}
\end{center}
\end{figure*}

In this section, we present FinLangNet, a dual-module framework designed to capture the distinct characteristics of heterogeneous financial data.  As illustrated in Figure \ref{fig:finlangnet}, our framework consists of two complementary branches: a \textbf{Non-Sequential Module} for static user profiles and a \textbf{Sequential Module} (SRG) for temporal behaviors. The outputs from both modules are fused into a unified embedding to generate risk predictions across multiple future time scales.

\subsection{Non-Sequential Module}
\paragraph{Motivation.} Non-sequential features (e.g., user profiles) contain complex static interactions that are crucial for defining baseline risk levels. Simple linear models often fail to capture the intricate combinations of these attributes (e.g., the combined risk of age, occupation, and income level).
\paragraph{Design.} We adopt DeepFM~\cite{guo2017deepfm} to capture complex interactions within the static feature vector $m \in \mathbb{R}^M$. DeepFM combines a Factorization Machine (FM) component for low-order (second-order) feature interactions and a deep neural network (DNN) component for high-order non-linear correlations. Concretely, the FM component is defined as
\begin{equation}
   y_{\text{FM}} = \langle w, m \rangle + \sum_{j_1=1}^{M} \sum_{j_2=j_1+1}^{M} \langle V_{j_1}, V_{j_2} \rangle m_{j_1} m_{j_2},
\end{equation}
where $w$ denotes linear weights and $V_j$ is the latent embedding for the $j$-th feature. In parallel, the DNN component takes the aggregated embedding signal and models higher-order interactions via an MLP:
\begin{equation}
\begin{split}
    y_{\text{DNN}} = \mathrm{MLP}\Bigg( \frac{1}{2} \bigg[ & \left( \sum_{i=1}^M m_i V_i \right)^2 
    - \sum_{i=1}^M \left( m_i^2 V_i^2 \right) \bigg] \Bigg).
\end{split}
\end{equation}
We combine the two parts to obtain the static interaction embedding:
\begin{equation}
O_{m} = f_{\text{DeepFM}}(m) = \sigma \left( y_{\text{FM}} + y_{\text{DNN}} \right),
\end{equation}
where $\sigma(\cdot)$ is a sigmoid function.
% \paragraph{Design.} We employ the DeepFM \cite{guo2017deepfm} architecture to model these interactions. Given static features $m \in \mathbb{R}^M$, the module simultaneously learns low-order interactions via a Factorization Machine (FM) component and high-order non-linear correlations via a Deep Neural Network (DNN) component.
% \begin{equation}
% O_{m} = f_{\text{DeepFM}}(m),
% \end{equation}
% where $O_m$ is the static interaction embedding. The detailed formulation of the FM and DNN components is provided in Appendix \ref{sec:appendix_method_deepfm}.

\subsection{Sequential Module (SRG)}
\paragraph{Motivation.} Financial sequences differ significantly from natural language: they are multi-source, highly sparse, and contain noise. Standard RNNs often struggle with long-term dependencies in such heterogeneous data. To address this, we propose the Sequence Representation Generator (SRG), which transforms continuous financial signals into discrete tokens and captures dependencies via a novel dual-prompt mechanism.

\paragraph{Discretization.} First, to mitigate data sparsity and robustness against noise, we discretize continuous features. For each channel $c$ in a source $z_s$, a tokenizer $\mathcal{T}_c$ transforms the continuous vector into discrete tokens $t_c \in \mathbb{N}^{T_s}$.

\paragraph{Dual-Prompt Mechanism.} 
Inspired by the `[CLS]` token in functional Transformers \cite{devlin2019bert}, we introduce prompts at two granularities to guide representation learning:
\begin{itemize}
    \item \textbf{Feature-level Prompt ($\widetilde{\phi}_c$):} Captures channel-specific global patterns. We prepend a learnable token to each channel sequence: $t_c' = (\widetilde{\phi}_c, t_{c,1}, \dots, t_{c,T_s})$. After embedding, the prompt $\widetilde{\phi}_c$ aggregates information unique to that feature dimension.
    \item \textbf{User-level Prompt ($P_s$):} Captures holistic user behavior across all channels. We aggregate channel-wise representations via weighted attention to form a unified sequence $H'$, and prepend a global learnable vector $P_s$.
\end{itemize}
The augmented sequence is processed by a Transformer encoder. The state of the user-level prompt at the final layer serves as the comprehensive sequential representation $O_{\text{ns}_s} = H^{(L)}[0]$. We concatenate representations from all $S$ sources to obtain $O_{\text{ns}}$.

\subsection{Multi-scale Credit Risk Prediction}
\paragraph{Motivation.} Credit risk is dynamic; a user might be safe in the short term but risky in the long term. A single prediction point is insufficient for comprehensive risk management. Furthermore, financial data suffers from severe class imbalance and varying sample difficulty.

\paragraph{Prediction \& Optimization.}
We fuse the static and sequential embeddings into a shared representation $O = [O_m; O_{\text{ns}}]$. To predict risks over $n$ different horizons (e.g., 30, 60, 90 days), we employ a multi-task setup where specific heads project $O$ to binary probabilities $\{y'_1, \dots, y'_n\}$. To robustly train this model, we design a {Dynamically Weighted Hybrid Loss}.

\paragraph{Weighted Logarithmic Loss (WLL).}
We assign higher penalties to the minority class (default cases):
\begin{equation}
\begin{split}
\mathcal{L}_{\text{WLL}, i} = - \bigg( & w^+ y_i \log(y'_i + \epsilon) \\
& + w^- (1 - y_i) \log(1 - y'_i + \epsilon) \bigg).
\end{split}
\end{equation}

\paragraph{Dynamic Hard Example Mining.}
We calculate a dynamic weight $\omega_i$ for each sample based on its prediction uncertainty, measured by the gradient norm $g_i = | \partial \mathcal{L}_i / \partial y'_i |$:
\begin{equation}
\omega_i = \frac{(g_i + \epsilon)^{-\alpha}}{\sum_{j} (g_j + \epsilon)^{-\alpha}}.
\end{equation}
This mechanism automatically up-weights samples where the model struggles.

\paragraph{Total Objective.}
The final objective balances regression (for probability smoothness) and classification stability:
\begin{equation}
\mathcal{L}_{\text{total}} = \frac{1}{n} \sum_{i=1}^{n} \omega_i \left[ \beta (y'_i - y_i)^2 + (1 - \beta) \mathcal{L}_{\text{WLL}, i} \right].
\end{equation}

\section{Experiment}
\begin{table*}[t]
\centering
\resizebox{\textwidth}{!}{
\begin{tabular}{lcccccccccccc}
\toprule
\multirow{2}{*}{\textbf{Model}} & \multicolumn{2}{c}{\textbf{$y_1(\tau=1)$}} & \multicolumn{2}{c}{\textbf{$y_2(\tau=2)$}} & \multicolumn{2}{c}{\textbf{$y_3(\tau=3)$}} & \multicolumn{2}{c}{\textbf{$y_4(\tau=4)$}} & \multicolumn{2}{c}{\textbf{$y_5(\tau=5)$}} & \multicolumn{2}{c}{\textbf{$y_6(\tau=6)$}} \\
\cmidrule(lr){2-3} \cmidrule(lr){4-5} \cmidrule(lr){6-7} \cmidrule(lr){8-9} \cmidrule(lr){10-11} \cmidrule(lr){12-13}
 & AUC & KS & AUC & KS & AUC & KS & AUC & KS & AUC & KS & AUC & KS \\
\midrule
\textit{Tabular \& Graph Baselines} & & & & & & & & & & & & \\
XGBoost & 72.78 & 32.85 & 75.76 & 37.42 & 70.89 & 30.00 & 73.04 & 33.18 & 68.69 & 26.96 & 69.70 & 28.31 \\
MLP \cite{goodfellow2016deep} & 71.97 & 31.81 & 74.76 & 36.14 & 69.95 & 28.70 & 72.00 & 31.76 & 67.59 & 25.38 & 68.45 & 26.55 \\
GraphSAGE \cite{hamilton2017inductive} & 72.19 & 32.04 & 75.09 & 36.49 & 70.26 & 29.09 & 72.37 & 32.16 & 68.03 & 26.00 & 68.95 & 27.25 \\
\midrule
\textit{Sequential Baselines} & & & & & & & & & & & & \\
GRU \cite{cho2014learning} & 72.59 & 32.40 & 75.68 & 37.16 & 70.93 & 30.05 & 73.37 & 33.57 & 69.06 & 27.44 & 70.62 & 29.75 \\
Transformer \cite{vaswani2017attention} & 72.54 & 32.62 & 75.95 & 37.98 & 70.97 & 30.12 & 73.76 & 34.54 & 69.30 & 27.82 & 71.19 & 30.67 \\
Mamba \cite{gu2023mamba} & 72.28 & 32.06 & 75.66 & 37.28 & 70.65 & 29.67 & 73.16 & 33.47 & 68.79 & 26.88 & 70.23 & 29.02 \\
TimesNet \cite{wu2023timesnet} & 72.49 & 32.54 & 75.90 & 37.98 & 70.83 & 29.99 & 73.48 & 34.15 & 69.26 & 27.73 & 71.05 & 30.53 \\
\midrule
\textit{LLM Baselines (Zero-Shot)} & & & & & & & & & & & & \\
DeepSeek-V3.2 \cite{deepseekai2025deepseekv3technicalreport} & 54.80  & 9.10  & 55.45  & 10.50  & 54.74 & 7.90 & 55.94 & 9.90 & 54.62 & 7.70 & 54.56 & 7.60 \\
GPT-4.1 \cite{gpt4.1} & 55.90  & 10.85  & 56.80  & 12.50  & 55.15 & 9.30 & 56.60 & 11.60 & 56.12 & 10.70 & 56.05 & 10.60 \\
\midrule
FinLangNet (Ours) & 73.55 & 34.08 & 76.96 & 39.46 & 71.92 & 31.60 & 74.51 & 35.67 & 70.33 & 29.27 & \textbf{72.12} & \textbf{32.16} \\
\textbf{FinLangNet + XGB (Fusion)} & \textbf{75.11} & \textbf{36.55} & \textbf{77.76} & \textbf{40.73} & \textbf{73.04} & \textbf{33.27} & \textbf{75.32} & \textbf{36.81} & \textbf{70.39} & \textbf{29.31} & {71.49} & {30.97} \\
\bottomrule
\end{tabular}
}
\caption{Performance comparison across multiple models and labels. The FinLangNet + XGB represents the deployed fusion strategy. The best results are highlighted in \textbf{bold}. All metrics are reported as percentages (\%).}
\label{tab:model_performance}
\end{table*}

\subsection{Experimental Setup}
We first evaluate our method on an industrial credit risk task and additionally verify it on public time series classification datasets.
\subsubsection{Industrial Credit Risk Assessment}

\noindent
\textbf{Dataset.} We evaluate FinLangNet on our large-scale proprietary industrial dataset comprising over 7 million users from a real-world financial service platform. The dataset includes 6 evaluation tasks representing overdue status predictions at different future time horizons (denoted as $\tau \in \{1, 2, ..., 6\}$), enabling assessment of both short-term and long-term risk forecasting capabilities. Dataset statistics are provided in Appendix Table~\ref{tab:Financial_Dataset}.

\noindent
\textbf{Evaluation Metrics.} Following industry standards for credit risk assessment, we employ two primary metrics: (1) the Kolmogorov-Smirnov (KS) statistic \cite{massey1951kolmogorov}, which measures the maximum divergence between cumulative distributions of positive and negative classes, directly quantifying discriminative power; and (2) Area Under the Curve (AUC), which evaluates the model's classification ability across all threshold settings.

\noindent
\textbf{Baselines.} We compare FinLangNet against a diverse set of methods: (1) XGBoost (the current production baseline); (2) deep sequential models including Transformer \cite{vaswani2017attention}, Mamba \cite{gu2023mamba}, and TimesNet \cite{wu2023timesnet}; and (3) state-of-the-art LLMs, specifically DeepSeek-V3.2 and GPT-4.1, evaluated in a zero-shot setting to assess general reasoning capabilities. Detailed information in Appendix~\ref{sec:app_hyper}.

\subsubsection{Public Time Series Benchmark}
\noindent
\textbf{Dataset.} To validate the generalizability of our sequential representation module (SRG), we evaluate on 5 multivariate time series classification (MTSC) tasks from the UEA archive \cite{bagnall2018uea}. These datasets span diverse domains including gesture recognition, audio analysis, and medical diagnostics, as detailed in Appendix Table~\ref{tb:UEA_Dataset}.

\noindent
\textbf{Evaluation Metrics.} Following standard practice for time series classification, we report classification accuracy as the primary metric.

\noindent
\textbf{Baselines.} We compare against 11 state-of-the-art MTSC methods including distance-based approaches (EDI, DTWI), deep learning models (MLSTM-FCNs \cite{karim2019multivariate}, TapNet \cite{zhang2020tapnet}, ShapeNet \cite{li2021shapenet}), feature-based methods (WEASEL-MUSE \cite{schafer2017multivariate}, Rocket \cite{dempster2020rocket}), and transformer architectures (TStamp \cite{zerveas2021transformer}, SVP-T \cite{zuo2023svp}).

\subsection{Results on Industrial Assessment}
\label{sec:exp_industrial}
Table \ref{tab:model_performance} presents the comparative evaluation on our large-scale industrial dataset across six temporal prediction tasks ($y_1$ to $y_6$). We compare FinLangNet against three categories of baselines: tabular methods, sequential deep learning, and Large Language Models. FinLangNet consistently outperforms all baselines, achieving significant improvements in both AUC and KS metrics. Specifically, compared to the strongest tabular baseline XGBoost, our method yields an average KS improvement of 6.3 pp, demonstrating the superiority of our multi-scale sequential learning approach over static feature engineering. Interestingly, while the fusion of FinLangNet and XGBoost generally achieves the state-of-the-art, it underperforms the standalone FinLangNet on the long-term task $y_6$. This suggests that XGBoost, which relies on static tabular features, struggles to capture long-range dependencies, thereby introducing noise rather than signal for distant prediction horizons.

Beyond raw performance, scalability is critical for industrial deployment. While some state-of-the-art methods like ShapeFormer \cite{le2024shapeformer} achieve high accuracy on small academic benchmarks, they rely on computationally expensive preprocessing (e.g., shapelet extraction from the entire dataset), making them intractable for our production environment with millions of users. In contrast, FinLangNet is designed for efficiency. Our dual-prompt mechanism processes multi-source sequences without heavy pre-computation, enabling real-time inference. This balance of high predictive accuracy and practical scalability confirms FinLangNet as a viable solution for large-scale industrial credit risk systems.
\subsection{Results on UEA Time Series Classification}
\begin{figure}[h]
  \centering
  \centerline{\includegraphics[width=0.95\linewidth]{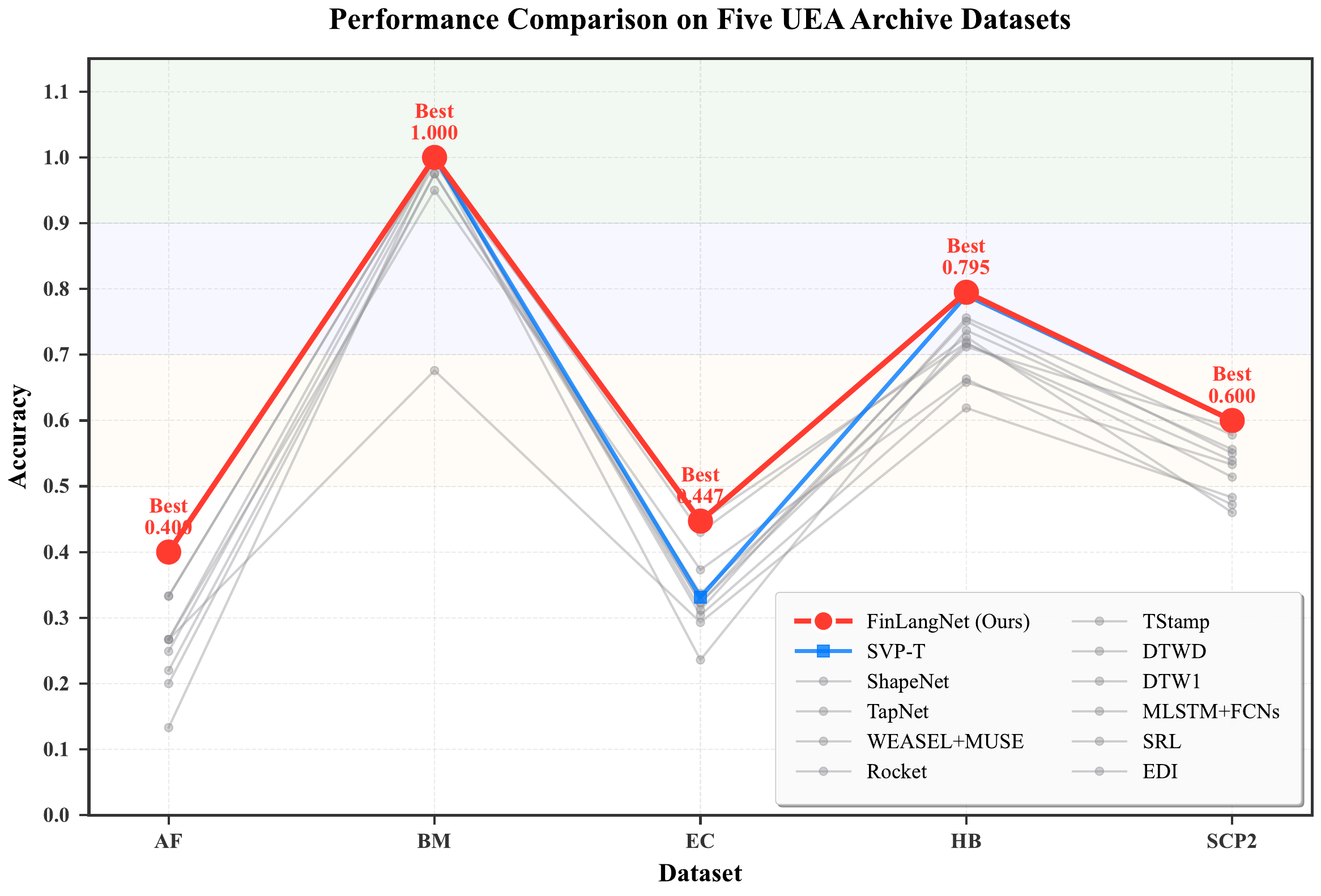}}
  \caption{Accuracy comparison on selected UEA datasets. Our SRG module demonstrates consistent superiority or competitiveness against SOTA benchmarks.} 
  \label{fig:uea_res} 
\end{figure}

To verify the generalization capability of our approach beyond financial domains, we evaluate the stand-alone SRG module on public UEA benchmarks. As illustrated in Figure \ref{fig:uea_res}, our method achieves competitive or state-of-the-art performance across diverse datasets ranging from medical diagnostics such as AtrialFibrillation to motion recognition like BasicMotions. Even without the full FinLangNet architecture by utilizing only the SRG module, our approach consistently outperforms transformer-based baselines such as TStamp and SVP-T. Detailed numerical comparisons are provided in Appendix Table \ref{tb:UEA Classification Results}.

These results highlight the versatility of our dual-prompt mechanism. While specifically designed to handle the heterogeneity of financial data, the SRG module proves effective in general multivariate time series classification tasks. Notably, it achieves this performance while maintaining the computational efficiency necessary for industrial deployment unlike heavy academic models like ShapeFormer. This confirms that the architectural innovations in FinLangNet, specifically the multi-granularity prompts, capture fundamental temporal dependencies applicable across various domains.

\begin{table*}[t]
  \centering
  \small
  \setlength{\tabcolsep}{4pt}
  \begin{tabular}{lc cccccccccccc c}
    \toprule
    \multirow{2}{*}{\textbf{Length Setup}} & 
    \multicolumn{2}{c}{\textbf{$y_1$}} &
    \multicolumn{2}{c}{\textbf{$y_2$}} &
    \multicolumn{2}{c}{\textbf{$y_3$}} &
    \multicolumn{2}{c}{\textbf{$y_4$}} & 
    \multicolumn{2}{c}{\textbf{$y_5$}} &
    \multicolumn{2}{c}{\textbf{$y_6$}} & \textbf{Avg KS} \\
    
    \cmidrule(lr){2-3} \cmidrule(lr){4-5} \cmidrule(lr){6-7} \cmidrule(lr){8-9} \cmidrule(lr){10-11} \cmidrule(lr){12-13} \cmidrule(lr){14-14}
    
     & AUC & KS & AUC & KS & AUC & KS & AUC & KS & AUC & KS & AUC & KS & (\%) \\
    \midrule
 
    1.00 $\times$ Length (Full) 
    & \textbf{73.55} & \textbf{34.08} & \textbf{76.96} & \textbf{39.46} & \textbf{71.92} & \textbf{31.60} & \textbf{74.51} & \textbf{35.67} & \textbf{70.33} & \textbf{29.27} & \textbf{72.12} & \textbf{32.16} & \textbf{33.71} \\
    
    0.50 $\times$ Length
    & 73.45 & 34.00 & 76.80 & 39.34 & 71.80 & 31.55 & 74.45 & 35.61 & 70.22 & 29.14 & 72.07 & 32.11 & 33.62 \\ 

    0.25 $\times$ Length
    & 73.36 & 33.82 & 76.77 & 39.18 & 71.72 & 31.30 & 74.37 & 35.54 & 70.14 & 29.07 & 71.99 & 32.00 & 33.49 \\ 
    \bottomrule
  \end{tabular}
   \caption{Performance robustness across varying historical sequence lengths. The model maintains high stability even when the input history is significantly truncated.}
  \label{tab:diff_length}
\end{table*}

\begin{table}[ht]
  \centering
  \small
  \begin{tabular}{lcc}
    \toprule
    \textbf{Configuration} & \textbf{KS (\%)} & \textbf{AUC (\%)}  \\
    \midrule
    \textbf{Baseline (No Prompts)} & 32.62 & 72.54 \\
    \midrule
    \textit{Prompt Contributions} & & \\
    + Feature Prompt (\( \widetilde{\phi}_c \)) & 32.79 & 72.65 \\
    + Source Prompt (\( P_s \)) & 32.99 & 72.78 \\
    \textbf{+ Both Prompts (Ours)} & \textbf{34.08} & \textbf{73.55} \\
    \midrule
    \textit{Module Importance} & & \\
    w/o Sequential Model (\(O_{\text{ns}} \)) & 12.53 & 58.62 \\
    w/o Non-sequential Model (\(O_m\)) & 32.75 & 72.66 \\
    \bottomrule
  \end{tabular}
  \caption{Ablation analysis of different modules on task $y_1(\tau = 1)$. The proposed dual-prompt mechanism ($ \widetilde{\phi}_c + P_s $) combined with sequential modeling achieves the best performance.}
  \label{tab:ablation_component}
\end{table}

\section{Ablation Study}

To understand the contribution of modules in FinLangNet, we conduct comprehensive ablation studies on the industrial credit risk dataset. All experiments strictly maintain consistent hyperparameters, varying only the component under evaluation.

\subsection{Component Analysis}

Table~\ref{tab:ablation_component} dissects the impact of different architectural choices. The results validate the efficacy of our dual-prompt mechanism where combining the feature-granularity prompt $\widetilde{\phi}_c$ and the data-source prompt $P_s$ yields the highest performance of 34.08\% KS. This outcome confirms that modeling both local feature nuances and global cross-source interactions provides complementary benefits. Crucially, removing the sequential module $O_{\text{ns}}$ leads to a catastrophic performance drop where the KS decreases by over 20\%, demonstrating that temporal behavioral dynamics are the dominant predictor of credit risk while static non-sequential features serve primarily as a supplementary signal.

\subsection{Impact of Sequence Length}

We further evaluate the model's robustness by truncating the historical sequence length while keeping non-sequential features fixed as shown in Table~\ref{tab:diff_length}. The performance remains remarkably stable even when the input sequence is reduced to only 25\% of its original length, with the average KS score dropping marginally from 33.71\% to 33.49\%. This results indicates that recent short-term behavioral patterns encode the most critical risk signals, which allows FinLangNet to generalize effectively to users with limited transaction histories and serves as a vital characteristic for solving the cold-start problem in production environments.

\subsection{Online A/B Test}
In real-world financial risk management, regulatory compliance bans the use of black-box models for direct credit decisioning. To bridge the gap between deep learning performance and industrial explainability, we designed a fusion framework. We deployed FinLangNet to abstract heterogeneous data streams (e.g., inquiry records, account ledgers, and behavioral logs) into a scalar language-risk sub-score, $s_{\text{lang}}$. This score is then integrated into our interpretability-centric XGBoost system as a dense feature, effectively enhancing the model's discriminative power without violating risk governance protocols. Theoretically complex but operationally efficient, the system runs on L20 GPU clusters, processing ~100 QPS with sub-100ms latency.

Online performance analysis detailed in Appendix~\ref{sec:appendix_deployment} and Figure~\ref{fig:abtest} demonstrates significant risk reduction. In a controlled A/B test at a representative 60\% approval threshold, the proposed framework reduced the default rate from 9.1\% in the benchmark to 8.2\%, which represents a 9.9\% relative improvement. This confirms that FinLangNet effectively discriminates risk by systematically rejecting high-risk applicants that traditional models miss. Further analysis reveals that combining sequential representations with domain features yields a 6.3 pp improvement in KS and a 12.3\% reduction in expected losses, thus validating the commercial value of the hybrid framework.

\section{Related Work}
\label{sec:app_related}
Credit risk prediction is a crucial task in the financial sector, typically focused on estimating the likelihood of borrower default over a specific time period. Credit scores, such as FICO \cite{maiden2024fico, lei2025zigong}, are widely used evaluation tools \cite{jensen1992using, bucker2022transparency}, generated by algorithms that analyze various user-related data to assess a borrower’s creditworthiness \cite{zhang2025timemaster}. Extensive research has explored the application of machine learning techniques for credit risk prediction, including decision-tree-based methods like XGBoost \cite{he2018novel},  graph models such as GraphSAGE \cite{balmaseda2023predicting}, ChebConv \cite{liu2022fast}, and their combinations \cite{fein2024single}. While deep learning methods are often considered to offer enhanced modeling capabilities, existing studies have found that XGBoost generally outperforms deep learning approaches in this domain \cite{xu2021loan}.

The processing of irregular, multi-source financial data is a critical challenge in credit risk prediction. Such data are typically presented in tabular form and often involve high-dimensional features, necessitating effective feature selection techniques. Several methods have been proposed to improve performance, including filter methods \cite{janane2023filter}, wrapper methods \cite{ahadzadeh2023sfe}, and embedded methods \cite{raghu2023sequential}, which enhance both model accuracy \cite{xu2024predict, ha2019improving, li2020heterogeneous} and interpretability \cite{ma2018study, xu2021loan}. Deep learning approaches have also been applied to handle these diverse data types \cite{gorishniy2021revisiting, borisov2022deep}, but they do not consistently outperform XGBoost models \cite{gorishniy2021revisiting} when working with tabular data.

From data perspective, another approach is to treat financial information as sequential data for processing. Structured data based on temporal sequences, such as transaction records or historical behaviors, can be represented in the standard time-series format, which encompass a wide range of methodologies. Models like EDI, DTWI, and DTWD \cite{bagnall2018uea} rely on calculating distances reflective of temporal warping or deviations in time sequences. MLSTM-FCNs \cite{karim2019multivariate} combine LSTM and CNN layers for feature generation, while WEASEL-MUSE \cite{schafer2017multivariate} transforms series into symbolic representations. In the domain of credit risk, time-series models include approaches from both statistical and machine learning \cite{el2022credit}, as well as deep learning methods \cite{forough2022sequential, ala2021modelling, wang2022deep, liang2023derisk} such as LSTM \cite{yu2019review} and Transformer \cite{vaswani2017attention, lei2025generative}. However, existing methods are predominantly built on preprocessed, well-structured financial time-series data, which limits their ability to handle irregular, multi-source records. Additionally, most approaches rely on relatively plain model architectures.

\section{Conclusion}

In this work, we present FinLangNet, a novel framework that addresses critical challenges in industrial credit risk assessment by treating it as a multi-scale behavioral prediction task. Our approach integrates heterogeneous data streams through a dual-architecture design that combines DeepFM for static features and the Sequential Representation Generation module with a dual-prompt mechanism for temporal data. Extensive experiments demonstrate FinLangNet's effectiveness with a 6.3 pp improvement in KS metric over XGBoost and a 9.9\% reduction in relative bad debt rate on industrial datasets while achieving state-of-the-art performance on UEA time series benchmarks. Successfully deployed in the international finance system of a leading ride-hailing platform serving millions of daily transactions, FinLangNet validates the feasibility of combining academic innovation with industrial practicality in high-stakes financial applications.

% \section*{Acknowledgments}

\bibliography{custom}

% \newpage
\clearpage

\appendix

\part*{Appendix}
The Appendix is organized as follows:
\begin{itemize}[leftmargin=1.5em, label={}]
    % \item \textbf{\ref{sec:app_related} Related Work} \\
    % \small Provides a comprehensive review of credit risk prediction, processing irregular multi-source financial data, and sequential modeling approaches.
    \item \textbf{\ref{sec:appendix_data} Data Details and Preprocessing} \\
    \small details the specifications of the large-scale Industrial Credit Risk Dataset (including non-sequential and sequential splits) and the public UEA benchmarks.
    % \item \textbf{\ref{sec:appendix_method} Methodology Details} \\
    % \small Presents the mathematical formulation of the DeepFM component.
    \item \textbf{\ref{sec:app_hyper} Hyperparameter Settings} \\
    \small Lists the detailed training configurations (Table \ref{tab:training_params}) and FinLangNet architecture specifications (Table \ref{tab:model_architecture}).
    \item \textbf{\ref{sec:app_uea_results} Extended UEA Benchmark Results} \\
    \small Validation of generalization capability on 5 UEA time-series datasets compared with state-of-the-art baselines.
    \item \textbf{\ref{sec:appendix_deployment} Real-World Deployment Analysis} \\
    \small An industrial case study, comparing FinLangNet with the production XGBoost model on thresholds, risk distribution, and operational impact.
\end{itemize}

\section{Data Details and Preprocessing}
\label{sec:appendix_data}

\begin{table*}[h]
  \vskip 0.1in
  \centering
  \small
  \renewcommand{\arraystretch}{1.0}
  \setlength{\tabcolsep}{4pt}
  \begin{tabular}{l|c|c|c|l}
    \toprule
    \textbf{Data Source} & \textbf{Channel} & \textbf{Seq. Len.} & \textbf{Dataset Split} & \textbf{Description} \\
    \midrule
     Basic Information & 11 & -- & \multirow{4}{*}{(3M, 0, 0.6M)} & Static Personal Profile\\
     Credit Report: Inquiry & 5 & 120 &  & External Credit Queries\\
     Credit Report: Account & 29 & 200 &  & History of Credit Lines\\
     Loan Behavior & 241 & 26 &  & Repayment \& Borrowing Logs\\
    \bottomrule
    \end{tabular}
     \caption{Statistics of the Industrial Credit Risk Dataset. \textbf{Channel} denotes the feature dimension ($C_s$), and \textbf{Seq. Len.} denotes the maximum sequence length ($T_s$).}
  \label{tab:Financial_Dataset}
\end{table*}

\begin{table*}[t]
\centering
\small
\setlength{\tabcolsep}{5mm}
\begin{tabular}{lccccc}
\toprule
\textbf{Dataset} & \textbf{Train Cases} & \textbf{Test Cases} & \textbf{Dimensions} & \textbf{Length} & \textbf{Classes}\\ 
\midrule
AtrialFibrillation (AF) & 15 & 15 & 2 & 640 & 3\\
BasicMotions (BM) & 40 & 40 & 4 & 100 & 4\\
EthanolConcentration (EC) & 261 & 263 & 3 & 1751 & 4\\
Heartbeat (HB) & 204 & 105 & 61 & 495 & 2\\
SelfRegulationSCP2 (SCP2) & 200 & 180 & 7 & 1152 & 2\\
\bottomrule
\end{tabular}
\caption{Details of the utilized UEA Multivariate Time Series Classification datasets.} 
\label{tb:UEA_Dataset}
\end{table*}

% Table 1: Training Parameters (FinLangNet)
\begin{table*}[thbp]
\centering
\small
\begin{tabular}{lll}
\hline
\textbf{Parameter}                  & \textbf{Value}   & \textbf{Description}                                                                                \\
\hline
Optimizer                           & AdamW            & Optimizer with weight decay for better generalization                                              \\
Learning Rate                       & 5e-4             & Initial learning rate with cosine annealing schedule                                                                    \\
Betas                               & (0.9, 0.999)     & Exponential decay rates for Adam moment estimates                                          \\
Epsilon                             & 1e-08            & Numerical stability constant for Adam                                                          \\
Weight Decay                        & 0.01             & L2 regularization coefficient                                                             \\
Loss Weight $\alpha$                & 0.5              & Weight for auxiliary task loss in multi-task learning                                                                \\
Loss Weight $\beta$                 & 0.5              & Balance between MSE (continuous) and WLL (categorical)                                                \\
Focal Weight $\gamma$               & 1.0              & Focal loss parameter for hard sample mining                                                                  \\
Training Epochs                     & 12               & Total training iterations over full dataset                                             \\
Batch Size                          & 512              & Number of samples per gradient update                                             \\
Gradient Clipping                   & 1.0              & Maximum gradient norm for stability                                             \\
Dropout Rate                        & 0.2              & Dropout probability for regularization                                             \\
Early Stopping Patience             & 3                & Epochs to wait before stopping if no improvement                                             \\
\hline
\end{tabular}
\caption{Training hyperparameters and optimization settings for FinLangNet.}
\label{tab:training_params}
\end{table*}

In this section, we provide comprehensive specifications for the datasets utilized in our experiments. We evaluate our method on two distinct types of data: (1) a large-scale industrial credit risk dataset, and (2) public benchmarks from the UEA Multivariate Time Series Archive.

\subsection{Industrial Credit Risk Dataset}
\begin{figure}[b]
    \centering
    \includegraphics[width=0.95\linewidth]{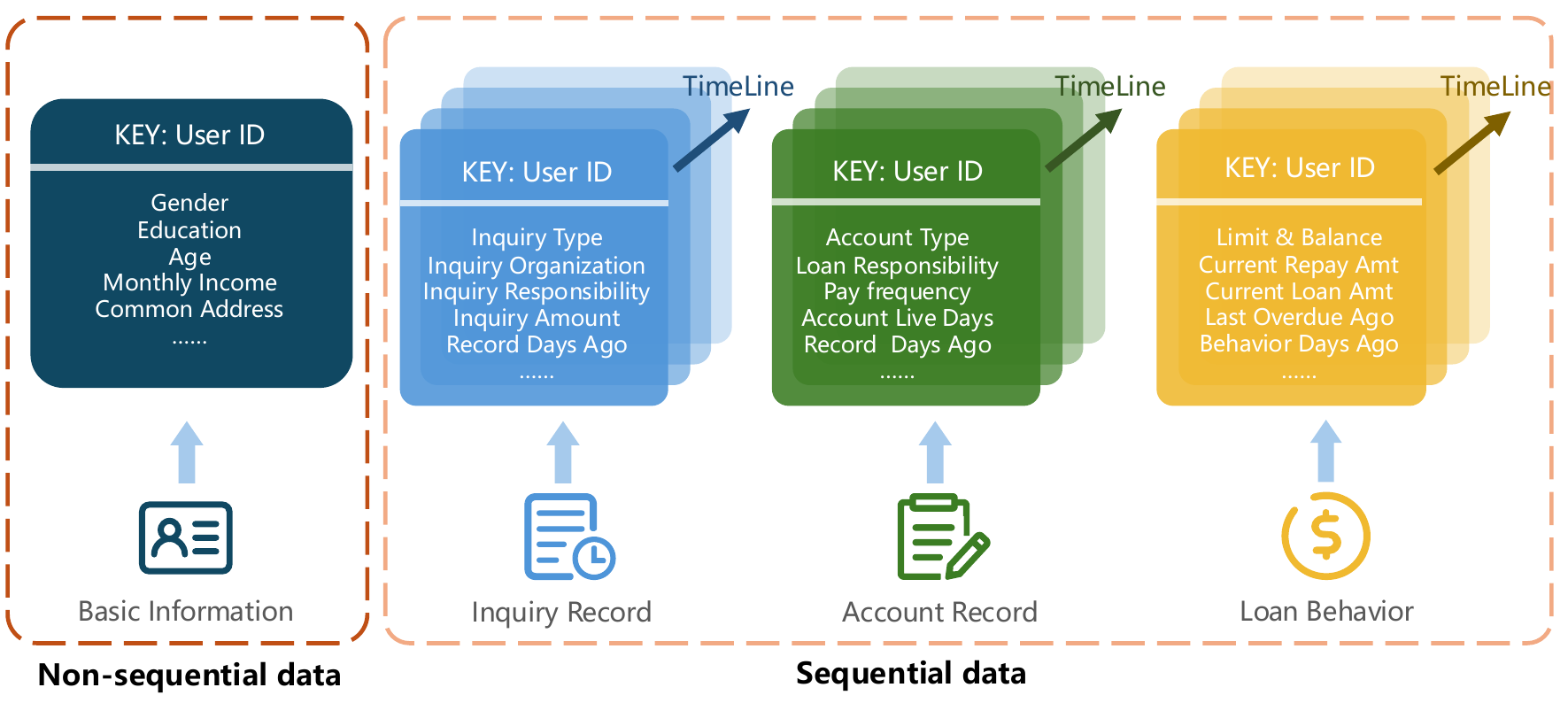}
    \caption{\small Structure of the Industrial Credit Risk Dataset. The input is categorized into Non-Sequential static data ($m$) and Sequential time-dependent data ($z$).}
    \label{fig:data_descri}
\end{figure}

We collected a real-world dataset from a leading international financial platform to evaluate credit risk prediction. The data originates from real business lines and is presented in tabular form. As illustrated in Figure \ref{fig:data_descri}, we reorganized the raw multi-source data into static and sequential components based on their temporal nature.

\paragraph{Data Composition.}
The input \(X\) is derived from three main parts:
i) \textbf{Basic Information}: Utilized as non-sequential data ($m$), containing invariant user attributes.
ii) \textbf{Credit Report}: Further split into Inquiry Records and Account Records. 
The reports are collected during the credit underwriting stage, 
with explicit user authorization granted to the platform for querying 
licensed third-party credit bureaus. 
These paid bureau inquiries provide detailed credit information 
at the individual record level.
iii) \textbf{Loan Behavior}: Logs of borrowing and repayment.
The latter two parts constitute the sequential data ($z$). We unified these records in strict chronological order. Statistical details are summarized in Table \ref{tab:Financial_Dataset}. The dataset involves approximately 3.6 million users, split temporally into 3 million for training and 0.6 million for testing.

\subsection{UEA Public Benchmarks}
To verify the generalization capability of our model, we also evaluate it on 5 representative multivariate time series classification datasets from the UEA Archive \cite{bagnall2018uea}. These datasets cover diverse domains including medical monitoring and human motion detection. The details are listed in Table \ref{tb:UEA_Dataset}.

% \section{Methodology Details}
\label{sec:appendix_method}

\paragraph{DeepFM Formulation}
\label{sec:appendix_method_deepfm}
In the Non-Sequential Module, we utilize DeepFM to capture interactions within the static feature vector $m \in \mathbb{R}^M$. 
The \textbf{FM component} models second-order feature interactions via inner products of latent vectors $V$:
\begin{equation}
   y_{\text{FM}} = \langle w, m \rangle + \sum_{j_1=1}^{M} \sum_{j_2=j_1+1}^{M} \langle V_{j_1}, V_{j_2} \rangle m_{j_1} m_{j_2}.
\end{equation}
Simultaneously, the \textbf{Deep component} models high-order interactions using a Multi-Layer Perceptron (MLP). The embeddings are fed into the network as:
\begin{equation}
\begin{split}
    y_{\text{DNN}} = \text{MLP}\Bigg( \frac{1}{2} \bigg[ & \left( \sum_{i=1}^M m_{i} V_{i} \right)^2 \\
    & - \sum_{i=1}^M \left( m_{i}^2 V_{i}^2 \right) \bigg] \Bigg).
\end{split}
\end{equation}

The final output combines both components: $O_{m} = \sigma(y_{\text{FM}} + y_{\text{DNN}})$.

% \subsection{Hybrid Loss with Dynamic Weighting}
% \label{sec:appendix_method_loss}
% To address class imbalance and focus on hard samples, we implement a specialized loss function.

% \paragraph{Weighted Logarithmic Loss (WLL).}
% We assign higher penalties to the minority class (default cases):
% \begin{equation}
% \begin{split}
% \mathcal{L}_{\text{WLL}, i} = - \bigg( & w^+ y_i \log(y'_i + \epsilon) \\
% & + w^- (1 - y_i) \log(1 - y'_i + \epsilon) \bigg).
% \end{split}
% \end{equation}

% \paragraph{Dynamic Hard Example Mining.}
% We calculate a dynamic weight $\omega_i$ for each sample based on its prediction uncertainty, measured by the gradient norm $g_i = | \partial \mathcal{L}_i / \partial y'_i |$:
% \begin{equation}
% \omega_i = \frac{(g_i + \epsilon)^{-\alpha}}{\sum_{j} (g_j + \epsilon)^{-\alpha}}.
% \end{equation}
% This mechanism automatically up-weights samples where the model struggles.

% \paragraph{Total Objective.}
% The final objective balances regression (for probability smoothness) and classification stability:
% \begin{equation}
% \mathcal{L}_{\text{total}} = \frac{1}{n} \sum_{i=1}^{n} \omega_i \left[ \beta (y'_i - y_i)^2 + (1 - \beta) \mathcal{L}_{\text{WLL}, i} \right].
% \end{equation}

\section{Hyperparameter Settings}
\label{sec:app_hyper}
We provide detailed hyperparameter configurations used in our experiments to ensure reproducibility. Table~\ref{tab:training_params} lists the optimization and training parameters for our proposed FinLangNet. Table~\ref{tab:model_architecture} details the specific architecture specifications of our model. Additionally, Table~\ref{tab:llm_settings} provides the specific versions and inference configurations for the Large Language Model (LLM) baselines.

% Table 2: Model Architecture (FinLangNet)
\begin{table*}[thbp]
\centering
\small
\begin{tabular}{lll}
\hline
\textbf{Component}                  & \textbf{Configuration}   & \textbf{Description}                                                                                \\
\hline
\multicolumn{3}{c}{\textit{Sequential Branch (SRG Module)}} \\
\hline
Embedding Dimension                 & 512              & Dimension of token embeddings                                              \\
Transformer Layers                  & 10                & Number of self-attention layers                                                                    \\
Attention Heads                     & 16                & Multi-head attention configuration                                          \\
FFN Hidden Size                     & 512              & Feed-forward network intermediate dimension                                                          \\
Position Encoding                   & Learnable        & Trainable positional embeddings                                                             \\
Max Sequence Length                 & 200              & Maximum tokens per sequence                                                                \\
Prompt Pool Size                    & 16               & Number of learnable soft prompts                                                \\
Prompt Length                       & 10               & Tokens per prompt template                                             \\
\hline
\multicolumn{3}{c}{\textit{Non-Sequential Branch (DeepFM)}} \\
\hline
FM Embedding Size                   & 16                & Factorization machine latent dimension                                             \\
DNN Hidden Layers                   & [512, 128, 64]   & Deep network architecture                                             \\
Activation Function                 & ReLU             & Non-linearity for hidden layers                                             \\
Batch Normalization                 & True             & Applied after each hidden layer                                             \\
\hline
\multicolumn{3}{c}{\textit{Fusion Layer}} \\
\hline
Fusion Method                       & Attention        & Cross-modal attention mechanism                                             \\
Output Dimension                   & 128              & Final representation size                                             \\
Temperature $\tau$                  & 0.07             & Contrastive learning temperature                                             \\
\hline
\end{tabular}
\caption{FinLangNet architecture specifications.}
\label{tab:model_architecture}
\end{table*}

% Table 3: LLM Baselines (New Added)
\begin{table*}[thbp]
\centering
\small
\begin{tabular}{lll}
\hline
\textbf{Model Family}               & \textbf{Configuration}   & \textbf{Description}                                                                                \\
\hline
\multicolumn{3}{c}{\textit{GPT-4 Series}} \\
\hline
Model Version                       & gpt-4.1-2025-04-14 & Specific snapshot API version utilized for experiments                             \\
Inference Temperature               & 0.0              & Set to zero for deterministic and reproducible evaluation                           \\
Top-p                               & 1.0              & Nucleus sampling probability                                                       \\
Max Output Tokens                   & 512              & Limit for reasoning chain and classification label                                  \\
\hline
\multicolumn{3}{c}{\textit{DeepSeek Series}} \\
\hline
Model Version                       & deepseek-v3-2-251201 & Specific snapshot version utilized for experiments                                \\
Inference Temperature               & 0.0              & deterministic generation setting                                                   \\
Repetition Penalty                  & 1.1              & Penalty to prevent circular reasoning in risk analysis                              \\
Context Window                      & 128k              & Full context window utilized for long behavior logs                                 \\
\hline
\multicolumn{3}{c}{\textit{Common Zero-Shot Setup}} \\
\hline
System Prompt                       & Expert Role      & "You are a senior credit risk expert..."                                           \\
Input Serialization                 & JSON             & Structured sequence representation for financial data                               \\
\hline
\end{tabular}
\caption{Configuration for Large Language Model (LLM) baselines. We utilize specific snapshot versions to ensure experimental consistency.}
\label{tab:llm_settings}
\end{table*}

\section{Extended UEA Benchmark Results}
\label{sec:app_uea_results}
To validate the generalization capability of FinLangNet beyond financial applications, we conducted comprehensive experiments on the UEA multivariate time series archive. Tables~\ref{tb:UEA Classification Results} present detailed comparisons with state-of-the-art time series classification methods.

\begin{table*}[thbp]
\scriptsize
\centerline{\setlength{\tabcolsep}{2mm}{
\begin{tabular}{lccccccccccccccc}
\toprule[1pt]

Data &EDI &DTWI &DTWD &MLSTM-FCNs &WEASEL+MUSE &SRL &TapNet &ShapeNet &Rocket &TStamp  &SVP-T  &FinLangNet\\ 
\midrule[1pt]

AF &0.267 &0.267 &0.220 &0.267 &0.333 &0.133 &0.333 &{\textbf{0.400}} &0.249 &0.200 &{\textbf{0.400}} &{\textbf{0.400}}\\
BM &0.676 &{\textbf{1.000}} &0.975 &0.950 &{\textbf{1.000}} &{\textbf{1.000}} &{\textbf{1.000}} &{\textbf{1.000}} &0.990 
&0.975 &{\textbf{1.000}} &{\textbf{1.000}}\\

EC &0.293 &0.304 &0.323 &0.373 &0.430 &0.236 &0.323 &0.312 &{\textbf{0.447}} &0.337 &0.331 &{\textbf{0.447}}\\
HB &0.619 &0.658 &0.717 &0.663 &0.727 &0.737 &0.751 &0.756 &0.718 &0.712 &0.790 &{\textbf{0.795}}\\

SCP2 &0.483 &0.533 &0.539 &0.472 &0.460 &0.556 &0.550 &0.578 &0.514 &0.589 &{\textbf{0.600}} &{\textbf{0.600}}\\

\bottomrule[1pt]
\end{tabular}
}}
\caption{Accuracies on Five Datasets of the UEA Archive} 
\label{tb:UEA Classification Results}
\end{table*}

\section{Real-World Deployment Analysis}
\label{sec:appendix_deployment}
\begin{figure*}[t]
\centering
\includegraphics[width=0.95\textwidth]{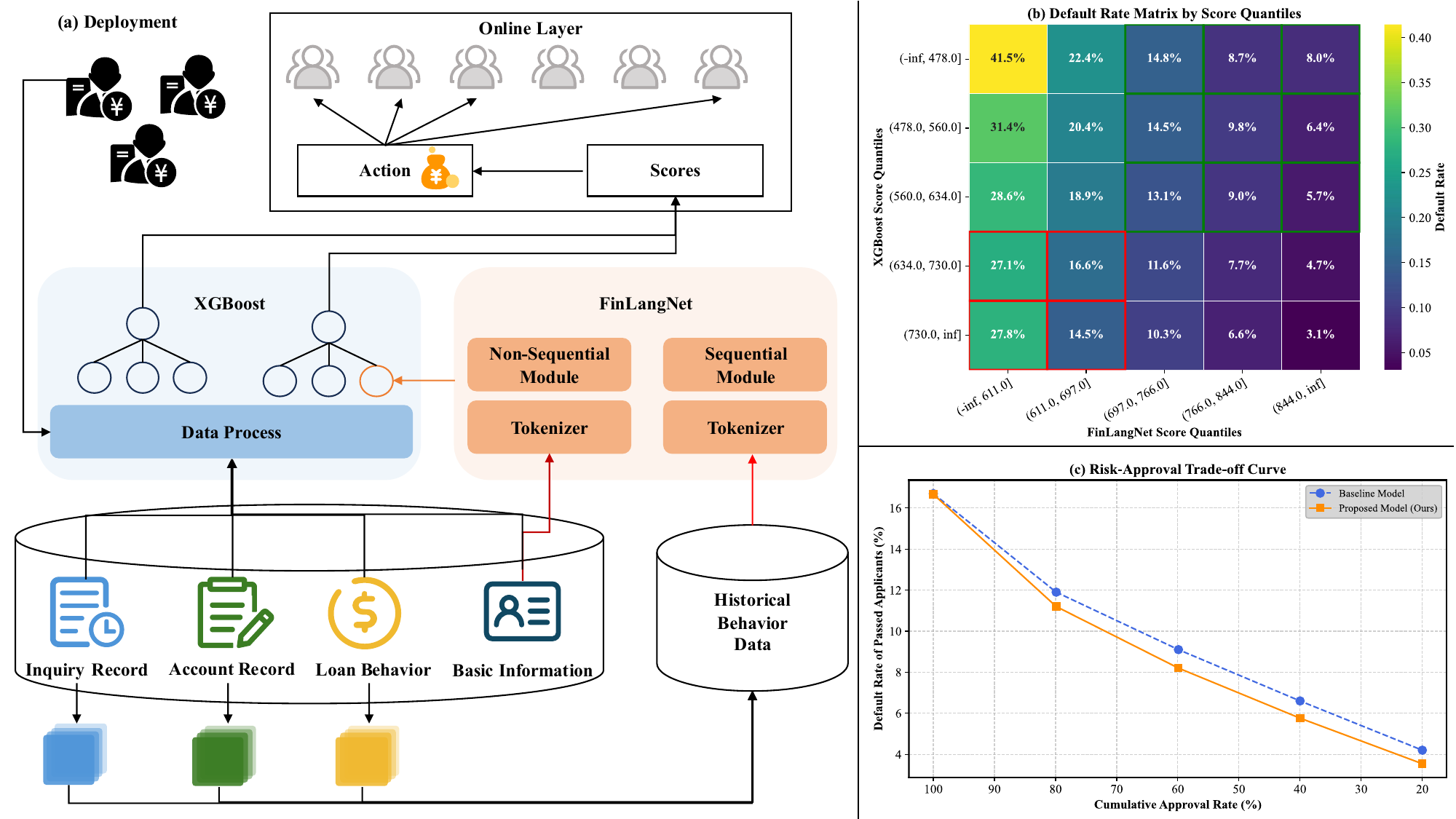}
\caption{\small Overview of Deployment. (a) Online deployment architecture and data flow. (b) Default‑rate matrix by score quantiles: The red-circled region indicates applicants rejected by FinLangNet but approved by the benchmark (risk reduction opportunities), while the green region shows potential viable customers. (c) Risk–approval trade‑off demonstrating consistent default rate reduction across varying approval thresholds. }
\label{fig:abtest}
\end{figure*}
To demonstrate the practical advantages of FinLangNet in industrial settings, we present a detailed case study comparing our approach with the production XGBoost model currently deployed at real-world . This analysis focuses on real-world performance metrics and operational considerations critical for risk management systems.
\subsection{Experimental Setup}
We compared two modeling paradigms:
\begin{itemize}[leftmargin=*,itemsep=0pt]
    \item \textbf{XGBoost (Baseline):} The production model utilizing 500+ manually engineered features derived from domain expertise, optimized over years of deployment.
    \item \textbf{FinLangNet:} Our proposed model processing raw sequential data without manual feature engineering, trained with multi-task learning across seven risk-related objectives.
\end{itemize}
For fair comparison, both models were evaluated on the common primary target $y_1$ (30-day delinquency, $\tau = 1$), which represents the most business-critical risk indicator. Performance was assessed on a held-out one-month test window following the training period.
\subsection{Threshold-Based Performance Analysis}
\begin{figure}[h]
    \centering
    \includegraphics[width=0.9\linewidth]{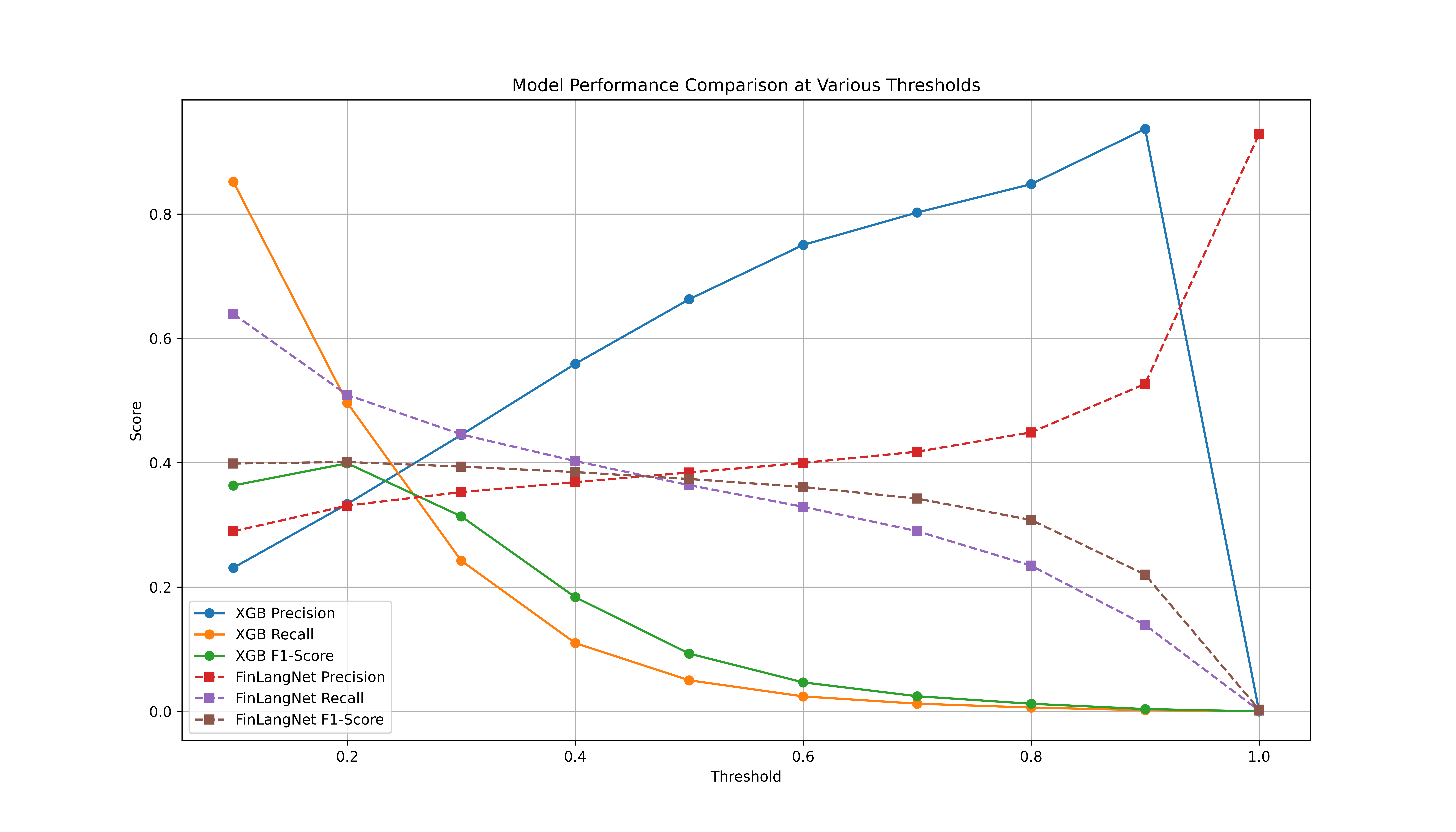}
    \caption{\small Performance comparison at various risk thresholds for $y_1(\tau = 1)$ prediction. FinLangNet demonstrates superior precision at operational thresholds (0.2--0.4) commonly used in production.}
    \label{fig:Thresholds}
\end{figure}

Risk control systems require careful threshold selection to balance between risk exposure (false negatives) and customer experience (false positives). Figure~\ref{fig:Thresholds} illustrates the precision-recall trade-offs at various decision thresholds.
Key observations include:
\begin{itemize}[leftmargin=*]
    \item \textbf{Low Thresholds (0.0--0.2):} XGBoost achieves higher recall but suffers from poor precision, creating operational challenges with excessive false positives.
    \item \textbf{Operational Range (0.2--0.4):} FinLangNet maintains balanced performance with significantly higher precision while preserving competitive recall, reducing manual review costs.
    \item \textbf{High Thresholds (0.4+):} Both models converge in performance, though FinLangNet maintains a slight precision advantage.
\end{itemize}
\subsection{Risk Distribution Analysis}
Figure~\ref{fig:scatterplot} reveals fundamental differences in how the models discriminate risk. 
\begin{figure}[h]
    \centering
    \includegraphics[width=0.9\linewidth]{Figs/com_xgb.png}
    \caption{\small Distribution of predicted risk scores versus actual labels. FinLangNet exhibits better separation between risk classes, particularly in the high-risk segment (scores > 0.6).}
    \label{fig:scatterplot}
\end{figure}
\begin{itemize}[leftmargin=*]
    \item \textbf{Risk Separation:} FinLangNet produces clearer separation between defaulters and non-defaulters, especially in the high-risk segment (predicted scores > 0.6).
    \item \textbf{Score Calibration:} While XGBoost shows some score clustering around certain values (likely due to dominant features), FinLangNet provides more continuous risk scoring.
    \item \textbf{False Positive Distribution:} FinLangNet's false positives tend to cluster in the moderate-risk range (0.3--0.5), making them easier to identify through secondary screening.
\end{itemize}
\subsection{Operational Impact Analysis}
Recognizing the complementary strengths of both approaches, we developed a hybrid strategy that incorporates FinLangNet's representations as additional features in the XGBoost framework. This integration leverages FinLangNet's automated feature learning from raw sequential data alongside XGBoost's domain-specific features.
The hybrid model achieved a \textbf{6.3 pp improvement in KS metric} over standalone XGBoost. To assess real-world deployment benefits, we analyzed swap sets—cases where the two models disagree on risk classification—and tracked their actual delinquency outcomes:
\begin{itemize}[leftmargin=*]
    \item \textbf{High-Risk Captures:} Among users classified as high-risk only by FinLangNet, 68\% showed delinquent behavior within 60 days, validating its superior risk detection.
    \item \textbf{False Positive Reduction:} FinLangNet correctly identified 42\% of XGBoost's false positives as low-risk, potentially reducing unnecessary credit restrictions.
    \item \textbf{Portfolio Performance:} Applying FinLangNet's risk scores would reduce the portfolio default rate by an estimated 12.3\% while maintaining the same approval rate.
\end{itemize}

\end{document}